\documentclass[aps,prb,showpacs,twocolumn,superscriptaddress,floatfix]{revtex4}

\usepackage{graphicx}
\usepackage{dcolumn}
\usepackage{bm}
\usepackage{amsmath}
\usepackage{amssymb}
\usepackage{color}

\begin{document}

\renewcommand{\vec}[1]{\mbox{\boldmath $#1$}}


\title{Spin-state responses to light impurity doping in low-spin perovskite LaCoO$_3$}


\author{Keisuke Tomiyasu}
\email[Electronic address: ]{tomiyasu@m.tohoku.ac.jp}
\affiliation{Department of Physics, Tohoku University, Aoba, Sendai 980-8578, Japan}
\author{Yuuki Kubota}
\affiliation{Department of Physics, Tohoku University, Aoba, Sendai 980-8578, Japan}
\author{Saya Shimomura}
\affiliation{Department of Physics, Tohoku University, Aoba, Sendai 980-8578, Japan}
\author{Mitsugi Onodera}
\affiliation{Department of Physics, Tohoku University, Aoba, Sendai 980-8578, Japan}
\author{Hironori Nakao}
\affiliation{Condensed Matter Research Center and Photon Factory, Institute of Materials Structure Science, High Energy Accelerator Research Organization, Tsukuba, Ibaraki 305-0801, Japan}
\affiliation{CREST, Japan Science and Technology Agency (JST), Tokyo, 102-0076 Japan}
\author{Youichi Murakami}
\affiliation{Condensed Matter Research Center and Photon Factory, Institute of Materials Structure Science, High Energy Accelerator Research Organization, Tsukuba, Ibaraki 305-0801, Japan}


\date{\today}

\begin{abstract}
We studied the spin-state responses to light impurity doping in low-spin perovskite LaCoO$_3$ (Co$^{3+}$: $d^6$) through magnetization and X-ray fluorescence measurements of single-crystal LaCo$_{0.99}M_{0.01}$O$_3$ ($M$ = Cr, Mn, Fe, Ni). In the magnetization curves measured at 1.8 K, a change in the spin-state was not observed for Cr, Mn, or Fe doping but was observed for Ni doping. Strong magnetic anisotropy along the [100] easy axis was also found in the Ni-doped sample. The fluorescence measurements revealed that the valences were roughly estimated to be Cr$^{3+}$, Mn$^{4+}$, Fe$^{(3+\delta)+}$, and Ni$^{3+}$. From the observed chemical trends, we propose that the chemical potential is a key factor in inducing the change of the low-spin state. By expanding a model of the ferromagnetic spin-state heptamer generated by hole doping [A. Podlesnyak {\it et al.}, Phys. Rev. Lett. {\bf 101}, 247603 (2008)], we discuss the emergence of highly anisotropic ferromagnetic spin-state clusters induced by low-spin Ni$^{3+}$ with Jahn--Teller activity. We also discuss applicability of the present results to mantle materials and impurity-doped pyrites with Fe ($d^6$).
\end{abstract}

\pacs{36.40.Cg, 75.30.Cr, 75.30.Wx, 75.47.Lx, 76.30.Fc, 91.35.-x}



\maketitle

\section{Introduction}
\begin{figure}[htbp]
\begin{center}
\includegraphics[width=3.0 in, keepaspectratio]{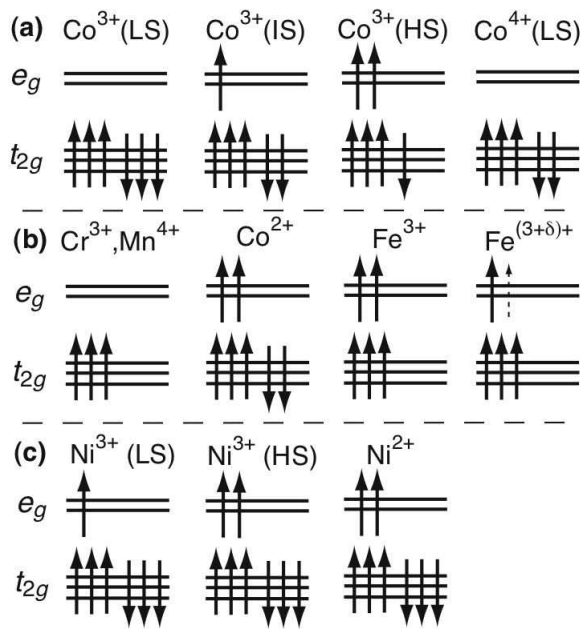}
\end{center}
\caption{\label{fig:el} Schematic diagrams of the $t_{2g}$ and $e_g$ orbital occupations for the 3$d$ transition metal ions in LaCoO$_3$. The up and down arrows indicate up and down spins, respectively. The dotted up arrow in Fe$^{(3+\delta)+}$ expresses partial occupation of either $e_g$ orbital corresponding to $\delta$.
}
\end{figure}

Transition metal oxides have been found to exhibit exotic electronic phenomena such as colossal magnetic resistivity and high-$T_c$ superconductivity, which are based on the strong correlations among the spin, orbital, charge, and lattice. In particular, perovskite-type cobalt oxide LaCoO$_3$ (Co$^{3+}$: $d^6$) is a rare inorganic material with spin-state degrees of freedom: low-spin (LS), intermediate-spin (IS), and high-spin (HS) states, as shown in Fig.~\ref{fig:el}(a).~\cite{Heikes_1964, Bhide_1972, Asai_1989, Sato_2009, Asai_1997} The Co$^{3+}$ ion is surrounded octahedrally by six O$^{2-}$ ions, and the crystal field splits the fivefold $d$ orbitals into triply degenerate $t_{2g}$ ($d_{xy}$, $d_{yz}$, $d_{zx}$) orbitals with a lower energy and doubly degenerate $e_g$ ($d_{3z^{2}-r^{2}}$, $d_{x^{2}-y^{2}}$) orbitals with a higher energy. The LS (HS) state appears when the crystal-field splitting energy is larger (smaller) than the Hund coupling energy, while the IS state is also theoretically possible through hybridization of the Co 3$d$ ($e_g$) and O 2$p$ orbitals when these energies are similar.~\cite{Korotin_1996}

LaCoO$_3$ exhibits insulating nonmagnetic LS states below about 20 K,~\cite{Heikes_1964, Bhide_1972, Asai_1989} and as the temperature increases, the LS states gradually change to homogeneous magnetic IS states or inhomogeneous states of LS and magnetic HS with semi-conductivity towards about 100 K. Further change to homogeneous HS states with metallic conductivity occurs around 600 K. LaCoO$_3$ therefore exhibits temperature-induced spin-state transitions.~\cite{Heikes_1964, Bhide_1972, Asai_1989} Such a spin-state transition can also be induced by a high magnetic field~\cite{Sato_2009} and by a high pressure applied to the material in the LS phase.~\cite{Asai_1997} Thus, these sensitive responses of the spin states demonstrate their near degeneracy in energy.

LaCoO$_3$ in an LS state also displays striking and fascinating responses to impurity doping. For example, very light Sr$^{2+}$ doping (below 1{\%}) of the La$^{3+}$ sites, which corresponds to hole doping from Co$^{3+}$ to Co$^{4+}$, generates a large magnetic moment of over 10 $\mu_{\rm B}$/hole, significantly larger than the maximum value of 5 $\mu_{\rm B}$ in $d$ electron systems.~\cite{Yamaguchi_1996} Further Sr doping leads to ferromagnetic spin glass, colossal magnetic resistivity, the anomalous Hall effect, and an insulator-to-metal transition.~\cite{Kriener_2004} In addition, Mn and Ni doping of the Co sites induces colossal magnetic resistivity, an insulator-to-metal transition, and ferromagnetic semi-conductivity with a high Curie temperature.~\cite{Jonker_1966, Hammer_2004} Both Mn- and Ni-doped LaCoO$_3$ are expected to be used in thin-film applications.~\cite{Robert_2008} Moreover, Co-site Fe doping induces weak ferromagnetism and a large magnetic anisotropy.~\cite{Karpinsky_2005, Karpinsky_2006, Karpinsky_2007}

The Sr-doping phenomena were proposed~\cite{Yamaguchi_1997, Louca_2003, Phelan_2006} and established~\cite{Podlesnyak_2008, Podlesnyak_2011} to microscopically originate from ferromagnetic spin-state polarons, which refer to spin clusters consisting of a core radical Co$^{4+}$ ion and its surrounding magnetic-spin-state Co$^{3+}$ ions changing from the LS state.~\cite{comm_1} The driving force is thought to be the double exchange interactions between Co$^{3+}$ and Co$^{4+}$; an electron is excited from the $t_{2g}$ to $e_g$ orbital in Co$^{3+}$ (LS to IS) and then hops between the Co$^{3+}$ and Co$^{4+}$ $e_g$ orbitals.~\cite{Yamaguchi_1997, Louca_2003} In inelastic neutron scattering studies for 0.2 to 1{\%} Sr doping, the best fit of the scattering vector ($\vec{Q}$) dependence of the intensities is provided by a spin heptamer model with a core radical Co$^{4+}$ ion and six nearest-neighbor Co$^{3+}$ ions (IS).~\cite{Podlesnyak_2008, Podlesnyak_2011} Pure isolated polarons are observed only for doping levels below 1{\%}, and the system enters the ferromagnetic spin glass phase owing to inter-polaron interactions at a doping level of at least 2{\%}.~\cite{Itoh_1994, Wu_2003, Podlesnyak_2011}

Based on the double exchange mechanism, Co-site doped LaCo$_{1-x}$$M_{x}$O$_3$ should also generate similar spin-state polarons when at least one majority-spin $e_g$ orbital is unfilled in the $M$ ions. Indeed, for $M=$ Ni$^{3+}$ (LS) with an unfilled $e_g$ orbital, powder magnetization measurements revealed that Co$^{3+}$ ions change to a magnetic spin state with ferromagnetic correlations at $x\geq0.01$.~\cite{Hammer_2004} For $M=$ Fe$^{3+}$ with filled $e_g$ orbitals, powder magnetization measurements, neutron diffraction, and M\"{o}ssbauer spectroscopy all indicated that the magnetic properties could be explained only by Fe$^{3+}$ magnetic moments without a spin-state change of Co$^{3+}$ at $x\geq0.15$.~\cite{Karpinsky_2005, Karpinsky_2006, Karpinsky_2007} However, for $M=$ Cr$^{3+}$ with an unfilled $e_g$ orbital, powder magnetization and neutron diffraction studies suggested that the Co$^{3+}$ LS state is maintained at $x\geq0.1$.~\cite{Tilset_1998} For $M=$ Mn with an unfilled $e_g$ orbital, X-ray absorption spectroscopy studies revealed that Mn doping generates a set of Mn$^{4+}$ and Co$^{2+}$ pair for each Mn ion (electron doping), and powder magnetization studies suggested that magnetism arises from Mn$^{4+}$ and Co$^{2+}$ without magnetic Co$^{3+}$.~\cite{Jonker_1966, Sikora_2006}

To explain this behavior and by analogy with the studies of very lightly hole-doped systems,~\cite{Yamaguchi_1996, Podlesnyak_2008} it is necessary to minimize the inter-dopant interactions and doping contamination in the system and to experimentally reinvestigate the perturbative influence of an isolated dopant embedded in the Co$^{3+}$ (LS) matrix. Thus, we performed a comparative study of single crystals of LaCo$_{0.99}$$M_{0.01}$O$_3$ ($M=$ Cr, Mn, Fe, Ni) with the same low concentration by conducting magnetization measurements complemented with X-ray fluorescence measurements, which have not been investigated thus far. Through these experiments we can determine whether the Co$^{3+}$ LS state changes for each $M$ atom, and for Ni doping, strong magnetic anisotropy is found for the first time. We discuss another key factor in the change of the Co$^{3+}$ LS state in addition to the double exchange mechanism. We also discuss applicability of the present results to mantle materials and impurity-doped pyrites with Fe ($d^6$).

\section{Experiments}
%
Single crystal rods approximately 6 mm in diameter of LaCoO$_3$ and LaCo$_{0.99}$$M_{0.01}$O$_3$ ($M=$ Cr, Mn, Fe, Ni) were grown in O$_2$ gas flow using the floating-zone method. La$_2$O$_3$, Co$_3$O$_4$, and $M_2$O$_3$ ($M=$ Cr, Mn, Fe, Ni) powders were used as the starting materials. In the powder specimens obtained by grinding a piece of the crystal, no reflections other than those of the perovskite structure were observed in the X-ray diffraction patterns. The composition ratios of the metal elements were evaluated by inductively coupled plasma (ICP); La:Co = 1:0.97 for the non-doped sample, La:Co:$M$ = 1:0.96:0.01 for the Cr-, Mn-, and Fe-doped samples, and 1:0.96:0.009 for the Ni-doped sample. The number of occupied $B$ sites decreased slightly but the total $B$-site occupation was constant at $0.97$ without any relative differences. In addition, the oxygen defect concentration $d$ defined by $AB$O$_{3-d}$ was evaluated by thermogravimetry: $d$ was $0.00\pm0.03$, $0.02\pm0.03$, $0.01\pm0.03$, $0.02\pm0.03$, and $0.01\pm0.03$ for the non-doped, Cr-, Mn-, Fe-, and Ni-doped samples, respectively, which are equivalent within the error range. Thus, these composition evaluations indicate that direct comparisons among these samples is meaningful.

Direct current magnetization measurements were performed with standard superconducting quantum interference device (SQUID) magnetometers, and crystal orientations were determined by X-ray Laue diffraction. Single crystal specimens of approximately 5 mm$^3$ were cut from the rods using a diamond disk cutter, fixed on a thin aluminum plate with varnish, and inserted into the magnetometers. X-ray fluorescence measurements were performed at room temperature on the BL-3A and BL-4C beamlines at the Photon Factory at KEK (Japan). To effectively extract the weak signals from the 1{\%} $M$ dopants, the incident X-ray energies were varied around the $K$-absorption edges of each $M$ atom, and the final energies were also fixed at each $K_{\alpha}$ energy with an energy-selective detector. Polycrystalline pellets of LaCrO$_3$, LaMnO$_3$, LaFeO$_3$, and (LaSr)NiO$_4$ with $M^{3+}$ synthesized by a standard solid-state reaction method were used for the reference samples.

\section{\label{sec:rslts}Results}
\begin{figure}[htbp]
\begin{center}
\includegraphics[width=3.1 in, keepaspectratio]{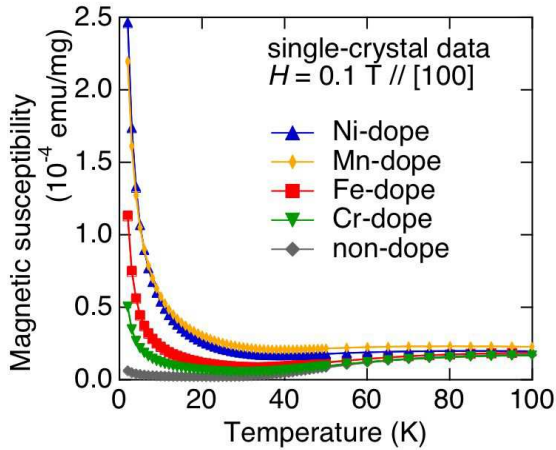}
\end{center}
\caption{\label{fig:MT} (Color online) Temperature dependence of the magnetic susceptibility in a magnetic field of 0.1 T applied along the [100] direction. The zero-field cooling and field cooling data with errors are of the same size as the symbols.}
\end{figure}

Figure~\ref{fig:MT} shows the temperature dependence of the magnetic susceptibility. All the doped samples exhibit much a larger increase in the susceptibility than the non-doped sample with decreasing temperature, indicating the appearance of a finite magnetic moment as a result of light doping. Furthermore, no difference between the zero-field-cooling and field-cooling curves was observed, indicating that there are no spin-glass-like components and that magnetic interactions among the dopants in the samples have been minimized.

\begin{figure*}[htbp]
\begin{center}
\includegraphics[width=0.90\linewidth, keepaspectratio]{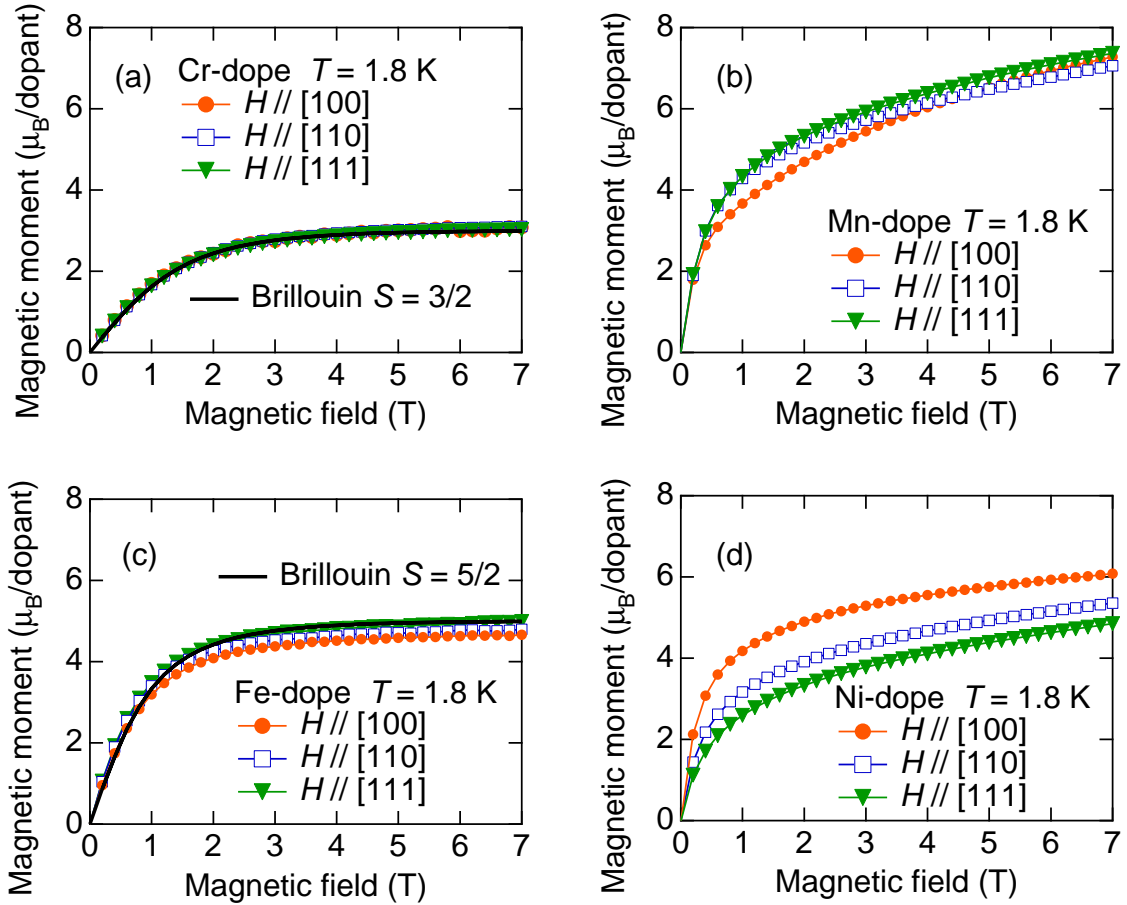}
\end{center}
\caption{\label{fig:MH} (Color online) Magnetization curves measured at 1.8 K with increasing magnetic field for the (a) Cr-, (b) Mn-, (c) Fe-, and (d) Ni-doped samples, respectively. To extract only the magnetization components generated by doping, the data of the non-doped sample were subtracted from the doped-sample data. The magnetic field was applied along the [100], [110], and [111] directions in all cases. In (a) and (c), Brillouin functions of $S=3/2$ and 5/2 are also shown without any fitting parameter. }
\end{figure*}
\begin{figure*}[htbp]
\begin{center}
\includegraphics[width=0.90\linewidth, keepaspectratio]{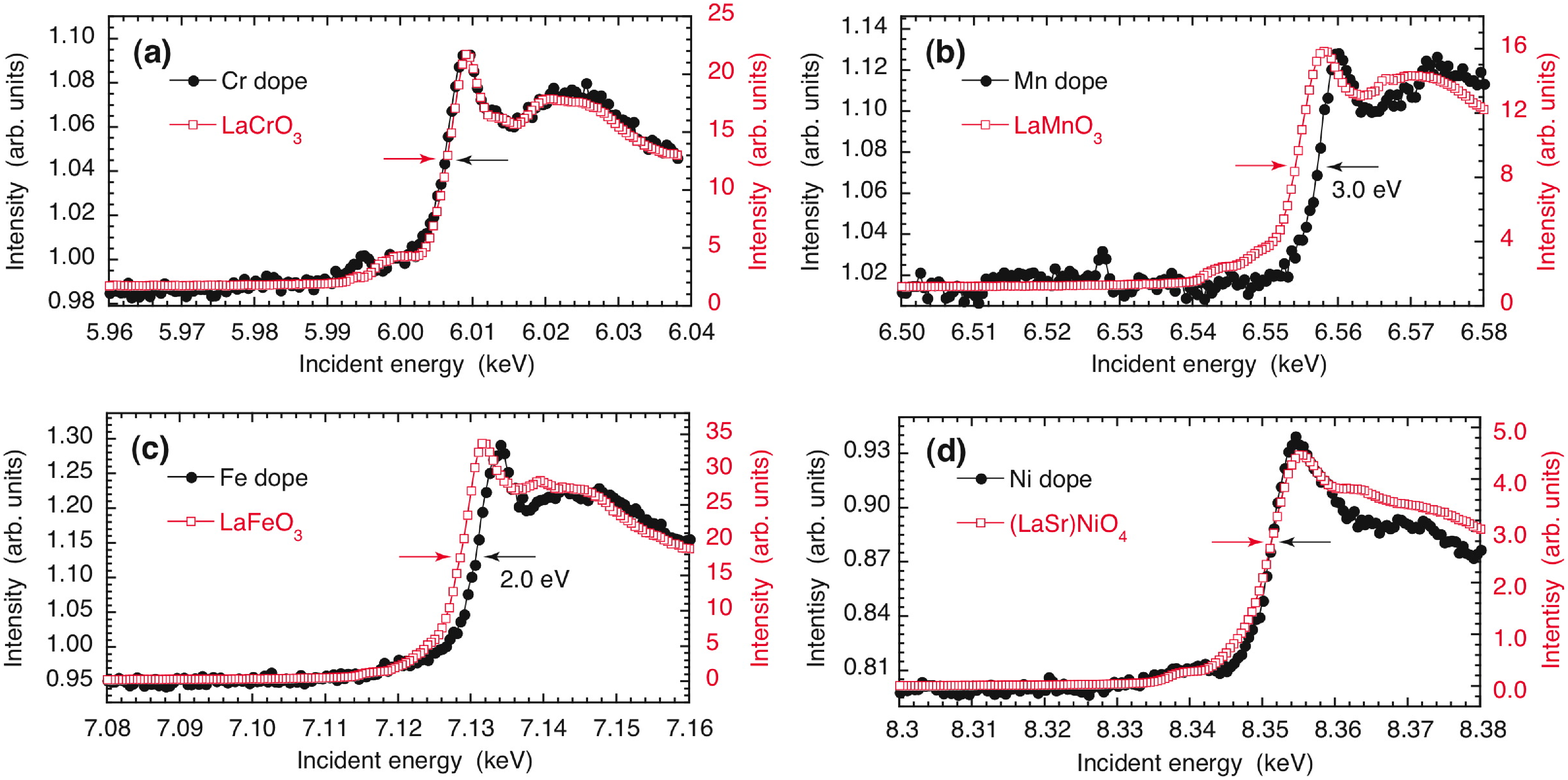}
\end{center}
\caption{\label{fig:PF} (Color online) X-ray fluorescence data measured at room temperature for the (a) Cr-, (b) Mn-, (c) Fe-, and (d) Ni-doped samples (solid black circles) compared with the data for the corresponding reference samples with their trivalent ions (open red squares), respectively. The arrow pairs denote the difference in the $K$ absorption edge positions defined as the inflection points. The lines are a guide for the eye. The statistical errors are smaller than the symbol sizes. }
\end{figure*}

Figures~\ref{fig:MH}(a)--\ref{fig:MH}(d) show the magnetization curves measured at 1.8 K. The Cr-doped sample curves are isotropic and follow the $S=3/2$ Brillouin function (Fig.~\ref{fig:MH}(a)). The Mn-doped sample curves are anisotropic along the [111] easy axis, and the magnetic moment is $\sim$7 $\mu_{\rm B}$ at 7 T (Fig.~\ref{fig:MH}(b)). The Fe-doped sample curves are slightly anisotropic along the [111] easy axis but mainly follow the $S=5/2$ Brillouin function (Fig.~\ref{fig:MH}(c)). Only the Ni-doped sample curves show a large anisotropy along the [100] easy axis (Fig.~\ref{fig:MH}(d)), and the magnetic moment is $\sim$6 $\mu_{\rm B}$ at 7 T. 

Figures~\ref{fig:PF}(a)--\ref{fig:PF}(d) show comparisons of the X-ray fluorescence spectra measured at room temperature of the 1{\%}-doped samples and the trivalent reference samples. In Fig.~\ref{fig:PF}(a), the $K$-absorption edge positions are nearly equal, indicating that the doped Cr ion is trivalent. In Fig.~\ref{fig:PF}(b), the edge position remarkably shifts to a higher energy relative to the reference edge, indicating that the Mn dopant ion is tetravalent, which is consistent with the X-ray absorption spectroscopy report for Mn doping of over 20{\%} in LaCoO$_3$.~\cite{Sikora_2006} In Fig.~\ref{fig:PF}(c), the edge position also shifts to a higher energy, indicating that the Fe dopant ion tends toward being tetravalent, Fe$^{(3+\delta)+}$. However, M\"{o}ssbauer spectroscopy determined the Fe valence to be trivalent for LaCo$_{0.42}$Fe$_{0.58}$O$_3$,~\cite{Karpinsky_2007} suggesting that this valence shift appears only at low Fe concentrations. In Fig.~\ref{fig:PF}(d), the edge positions are nearly equal, and so the Ni dopant ion is roughly trivalent.

\section{Discussion}
\subsection{\label{sec:CrMnFe}Cr, Mn, and Fe doping}
%
The results of the magnetization and fluorescence measurements demonstrate that the nonmagnetic LS state of Co$^{3+}$ is robust against Cr$^{3+}$ doping.

For Mn doping, since Mn$^{4+}$ is equivalent to Cr$^{3+}$ in terms of $d^3$, the Mn$^{4+}$ ion is expected to exhibit similar isotropic magnetization curves to those of Cr$^{3+}$. The paired Co$^{2+}$ ($d^7$: $S=3/2$, effective orbital angular momentum $l=1$) is expected to exhibit a total moment of $\sim$4 $\mu_{\rm B}$ with strong magnetic anisotropy owing to the $\vec{S}$ and $\vec{l}$ combination, as in CoO,~\cite{Singer_1956} to which the observed anisotropy is attributed. 
Further, the summation of Mn$^{4+}$ and Co$^{2+}$ alone gives a moment of $\sim$7 $\mu_{\rm B}$, which is in agreement with the present experimental value at 7 T (Fig.~\ref{fig:MH}(b)). All these facts can be explained by the Mn$^{4+}$ and Co$^{2+}$ alone without the magnetic state of Co$^{3+}$, indicating that the Co$^{3+}$ matrix mainly keeps the LS state. 
However, the magnetization is not yet saturated at 7 T unlike that of the Cr-doped sample, suggesting that the Mn doping accompanied with generation of Co$^{2+}$ has a subtle but finite effect on the Co$^{3+}$ matrix. We discuss this point in terms of an ionic radius in Sec.~\ref{sec:origin}. 

The magnetization curves for the Fe-doped sample were close to the $S=5/2$ Brillouin function, and no giant magnetic moment was observed, unlike that in the Sr-doped system. Therefore, the Fe ion will be effectively magnetically isolated with $S=5/2$ and the Co$^{3+}$ LS state will be maintained. Further, unfilled $e_g$ orbital often leads to an instantaneously unfilled $t_{2g}$ orbital state through a second perturbative excitation and de-excitation process, which is accompanied by the $\vec{l}$.~\cite{Ohtani_2010, Sagayama_2011} Thus, the observed anisotropy is again attributed to the $\vec{l}$ of Fe$^{(3+\delta)+}$ ($d^{5-\delta}$, $S=(5-\delta)/2$) with unfilled $e_g$ orbital (Fig.~\ref{fig:el}). Strictly speaking, however, the $S=(5-\delta)/2$ is inconsistent with the value of 5 $\mu_{\rm B}$ observed at 7 T. To explain the inconsistency, we consider that the $\delta$ electron lost from Fe expands to the surrounding O$^{2-}$ and Co$^{3+}$ by catalytically borrowing their orbitals, in a similar manner to what occurs in a complex, but is still ferromagnetically coupled with the Fe spins, which behave magnetically as $S=5/2$. However, the true microscopic picture of the lost electron is beyond the scope of this study.

\subsection{\label{sec:Ni}Ni doping}
%
For the Ni-doped sample, the exact electronic state has been somewhat controversial so far; it could be trivalent or divalent and may depend on the Ni concentration and subtle sample synthesis conditions.~\cite{Rao_1975, Kobayashi_1999, Hammer_2004, Ivanova_2010, Yu_2010} Thus, we discuss here three states: Ni$^{3+}$ (LS), Ni$^{3+}$ (HS), and the Ni$^{2+}$ and Co$^{4+}$ pair. First, Ni$^{3+}$ (LS) is Jahn--Teller active with $e_{g}$ orbital symmetry, which agrees with the observed [100] magnetic anisotropy. Second, in contrast to this, since Ni$^{3+}$ (HS) is equivalent to Co$^{2+}$ in terms of $d^7$ with $\vec{l}$, [111] anisotropy like that seen in the Mn- and Fe-doped sample magnetizations is expected, which is inconsistent with the present experimental results. Third, Ni$^{2+}$ is isotropic without either Jahn--Teller activity or $\vec{l}$, and the Co$^{4+}$ should exhibit a giant magnetic moment of over 10 $\mu_{\rm B}$. Both of these are inconsistent with the experimental results. Thus, we conclude that the doped Ni ion takes the trivalent LS state.

\begin{figure}[htbp]
\begin{center}
\includegraphics[width=3.1 in, keepaspectratio]{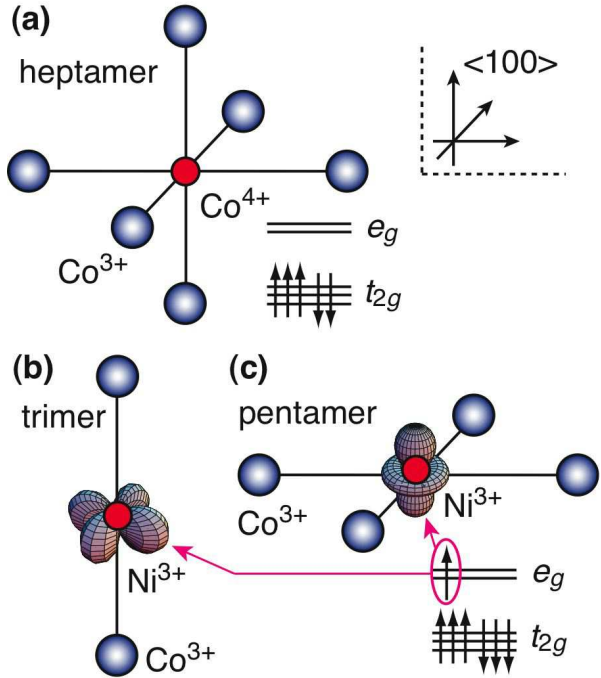}
\end{center}
\caption{\label{fig:model} (Color online) Molecular models of the ferromagnetic spin-state clusters. (a) Heptamer model for hole doping (Co$^{4+}$).~\cite{Podlesnyak_2008} (b) Trimer and (c) pentamer models for LS Ni$^{3+}$ doping with one $d_{x^{2}-y^{2}}$ electron and one $d_{3z^{2}-r^{2}}$ electron, respectively. The occupied $e_g$ orbitals are also depicted at the Ni positions. }
\end{figure}

On the basis of this conclusion, since Ni$^{3+}$ (LS) exhibits a moment of only 1 $\mu_{\rm B}$ (Fig.~\ref{fig:el}), which is far below the experimental value of $\sim$6 $\mu_{\rm B}$ at 7 T, its surrounding Co$^{3+}$ ions change from LS to another magnetic spin state. Interestingly, in Ni$^{3+}$ (LS), one majority-spin $e_g$ orbital is empty, unlike in Co$^{4+}$ in which both majority-spin $e_g$ orbitals are empty (hole doping). Therefore, based on the double exchange mechanism,~\cite{Podlesnyak_2008} we can propose the emergence of anisotropic ferromagnetic spin-state clusters, that is, a bar magnet for the $d_{3z^{2}-r^{2}}$ empty orbital and a plane magnet for the $d_{x^{2}-y^{2}}$ empty orbital, as shown in Figs.~\ref{fig:model}(b) and \ref{fig:model}(c), respectively.

At this stage, however, the magnetic Co$^{3+}$ cannot be identified as either IS, HS, or another possible spin state with non-integer $e_g$ electrons. Also, the electronic states might be unique to the light doping and are not necessarily maintained at higher Ni concentrations. These points should be studied in the future.

\subsection{\label{sec:origin}Origin of the spin-state change in Co$^{3+}$}
%
We thus now seek an answer to the question ``what makes the Co$^{3+}$ LS state change?" The present experimental results revealed that this change does not occur almost completely for Cr or Fe doping, mainly for Mn or Co$^{2+}$ (electron doping by Mn substitution) doping, but does occur for Co$^{4+}$ (hole doping by Sr substitution)~\cite{Yamaguchi_1996} and Ni doping. Interestingly, we note that Cr, Mn, and Fe are all lower group elements compared to Co and that Ni is in a higher group. This strongly suggests that a key factor is the chemical potential energy (Fermi energy or atomic orbital energy) of an outermost electron. For these elements, the energy is ordered as Cr $>$ Mn $>$ Fe $>$ Co$^{2+}$ $>$ Co$^{3+}$ $>$ Co$^{4+}$ $>$ Ni with a $\sim$1 eV difference,~\cite{Latter_1955, Ishii_2009} where the order of the Co ions is based on the Coulomb potentials. Thus, in the spin-state change, instead of losing excitation energy in changing from the $t_{2g}$ to $e_{g}$ orbital in Co$^{3+}$, an excited $e_{g}$-electron cloud flows slightly from the Co$^{3+}$ to the dopant ion so as to lower the total energy not only through the kinetic double exchange (normally $<\sim$0.1 eV) but also through the lower chemical potential of the dopant ($\sim$1 eV). 

The lattice strain and chemical pressure will also have an effect on the LS state change; magnetic spin (LS) states favor tensile (compressive) strain, as suggested by the existence of a pressure-induced spin-state transition.~\cite{Asai_1997} Density functional calculations have also suggested that LaCoO$_3$ exhibits a HS/LS mixed state for tensile strain over the critical magnitude, while the LS state is robust for compressive strain.~\cite{Seo_2012} Thus, we here evaluate the critical minimum ionic radius $r_{\rm min}$ of dopants. Table~\ref{tab:r} summarizes the ionic radii $r$ of the dopants and matrix Co$^{3+}$ in the literature.~\cite{Shannon_1976} First, nonmagnetic Al and Ga doping maintains the LS state even at high concentration.~\cite{Kyomen_2003} This Al ineffectiveness is consistent with the robustness of LS state for compressive strain, whereas the Ga ineffectiveness indicates $r_{\rm Ga^{3+}}<r_{\rm min}$. 

Second, nonmagnetic Rh doping barely maintains the LS state but exhibits an appreciably larger increase in magnetic susceptibility than the non-doped system at a doping level of only 2{\%}.~\cite{Knizek_2012} Above 4{\%}, the HS state with ferromagnetic correlations appears,~\cite{Kyomen_2003, Asai_2011, Knizek_2012} which was ascribed to the large $r_{\rm Rh^{3+}}$ in other density functional calculations.~\cite{Knizek_2012} Thus, $r_{\rm Rh^{3+}}$ is thought to exceed $r_{\rm min}$. 
Third, in the present Mn-doped sample magnetization data also, the absence of saturation suggests that the LS state is barely maintained but suffers a subtle effect, as mentioned in Sec.~\ref{sec:CrMnFe}. Further, other magnetization measurements detected slight sign of ferromagnetic correlations at a Mn doping level of 5{\%},~\cite{Autret_2005} which is also similar to those in the Rh-doped system. Thus, the subtle effect is also attributed to strain and $r_{\rm Mn}$ will be beyond $r_{\rm min}$. 
In this way, we obtain $r_{\rm min}\simeq0.63\pm0.01$ {\AA} as a value between $r_{\rm Ga^{3+}}$ and $r_{\rm Mn}$. This value is obviously larger than $r_{\rm Co^{3+} (HS)}$, which is in agreement with the HS state appearance model. 

\begin{table}[htbp]
\caption{\label{tab:r} Shannon ionic radii in descending order in {\AA} units.~\cite{Shannon_1976} The average value of $r_{\rm Mn^{4+}}$ (0.53 {\AA}) and $r_{\rm Co^{2+}}$ (0.745 {\AA}) for Mn doping and $r_{\rm Fe^{4+}}$ (0.585 {\AA}) to $r_{\rm Fe^{3+}}$ (0.645 {\AA}) for Fe doping are listed. }
\begin{ruledtabular}
\begin{tabular}{cc|cc}
Ion & Radius & Ion & Radius \\
\hline
$r_{\rm Rh^{3+}}$ & 0.665 & $r_{\rm Co^{3+} (HS)}$ & 0.61\\
$r_{\rm Mn}$ & 0.638 & $r_{\rm Ni^{3+} (LS)}$ & 0.56\\
$r_{\rm Ga^{3+}}$ & 0.62 & $r_{\rm Co^{3+} (LS)}$ & 0.545\\
$r_{\rm Cr^{3+}}$ & 0.615 & $r_{\rm Al^{3+}}$ & 0.535\\
$r_{\rm Fe^{(3+\delta)+}}$ & $0.615\pm0.03$\\
\end{tabular}
\end{ruledtabular}
\end{table}
%

%
\subsection{Applications of the present results}
We discuss applicability of the present results on the $d^6$ LS state change. First, the spin-state responses to Cr, Mn, Fe, Ni, Al, Ga, and Rh doping are summarized as follows. Ni doping will sensitively induce magnetic Co$^{3+}$ via the chemical potential combined with double exchange. Rh and Mn doping will gradually induce magnetic Co$^{3+}$ via strain in accordance with a doping level and magnitude of magnetic field. All other doping with neither a low chemical potential nor a large ionic radius will maintain the LS state of Co$^{3+}$. 

The next example of the application is impurity effects in the Earth's most abundant phase, perovskite (Mg, Fe)(Si, Al)O$_3$. This Fe atoms take divalent $d^6$ states, and change from HS towards LS states with increasing the mantle depth and pressure.~\cite{McCammon_2008} Interestingly, this spin-state change is suppressed in lightly Al doped (Mg$_{0.86}$Fe$_{0.14}$)(Si$_{0.98}$Al$_{0.02}$)O$_{3}$ compared to in (Mg$_{0.88}$Fe$_{0.12}$)SiO$_{3}$.~\cite{McCammon_2008} Although the physical origin has not been clarified thus far probably because of experimental difficulty under extremely high pressure, the present results provide the following interpretation. The Al$^{3+}$ substitution of Si$^{4+}$ corresponds to the hole doping to Fe sites like the Sr$^{2+}$ substitution of La$^{3+}$. The Fe$^{2+}$ LS state change will be therefore significant via the chemical potential with approaching the LS state, which is attributable to the origin. Furthermore, Al$^{3+}$ is much larger than Si$^{4+}$ (0.400 {\AA}),~\cite{Shannon_1976} which will  be also another origin. 

The last example is impurity effects in semi-conductive pyrite FeS$_2$, which are difficult to understand in the itinerant picture. This material is positioned between localization and itinerancy, and Fe $d$ bands can be described by the same LS state ($d^6$: $(t_{2g})^6$, $(e_{g})^0$, 0 $\mu_{\rm B}$).~\cite{Folkerts_1987} Although magnetic doping to Fe sites is supposed to generate the RKKY screening, the matrix Fe LS state does not change for Mn ($d^5$: $(t_{2g})^3$, $(e_{g})^2$, 5 $\mu_{\rm B}$) or Ni ($d^8$: $(t_{2g})^6$, $(e_{g})^2$, 2 $\mu_{\rm B}$) doping~\cite{Adachi_1971, Adachi_1976} but does change for Co ($d^7$: $(t_{2g})^6$, $(e_{g})^1$, 1 $\mu_{\rm B}$) doping.~\cite{Guo_2008} The Fe$_{1-x}$Co$_{x}$S$_{2}$ enters a ferromagnetic spin-glass phase at a doping level of only 1 {\%}.~\cite{Guo_2008} On the other hand, these chemical trends are consistent with the present results. The Mn-doped FeS$_2$ is magnetically nearly equivalent to the Fe-doped LaCoO$_3$ keeping the Co$^{3+}$ LS state (Fe$^{(3+\delta)+}$: nearly $d^5$ and 5 $\mu_{\rm B}$). In Ni-doped FeS$_2$, since all majority-spin orbitals are filled in the Ni, which forbids the double exchange, the Fe LS state will be maintained. In contrast, in terms of the sensitive appearance of ferromagnetic correlations, the Co-doped FeS$_2$ coincides with the Ni-doped LaCoO$_3$ with ferromagnetic spin-state clusters. 

Despite the same doping level of 1{\%}, however, the insulating Ni-doped LaCoO$_3$ is not yet a spin glass unlike the semi-conductive Co-doped FeS$_2$. The latter spin-glass nature is thought to be due to the RKKY screening.~\cite{Guo_2008} Thus, this difference is probably a limitation of applicability of the present results, which are based on the localized picture. 

\section{Summary}
In summary, we have studied the impurity-induced spin-state responses in lightly doped single crystals of LaCo$_{0.99}M_{0.01}$O$_{3}$ ($M$ = Cr, Mn, Fe, Ni) using magnetization and X-ray fluorescence measurements. A distinct spin-state change from the LS state in Co$^{3+}$ was observed only in the Ni-doped sample, which was accompanied by strong magnetic anisotropy. This change, which has been demonstrated experimentally for the first time, suggests the emergence of {\it anisotropic} ferromagnetic spin-state clusters as the expanded model of a spin-state heptamer generated by hole doping. Further, the results of the present experiments suggest that the chemical potential is also a key factor in the spin-state transition, in addition to the temperature, magnetic field, pressure (strain), and double exchange mechanism. For the strain-driven spin-state responses, the critical minimum ionic radius of a lightly doped impurity was evaluated. Taking for example mantle materials and impurity-doped pyrites with Fe, which has the same $d^6$ spin-state degree of freedom as the Co$^{3+}$ in LaCoO$_3$, we also discussed applicability of the present results. Future studies of the spin-state degree of freedom could develop interdisciplinary physics. 

\acknowledgments 
We are grateful to Mr. S. Kayamori of the Department of Instrumental Analysis, Technical Division, School of Engineering at Tohoku University for supporting the ICP analysis, and to the staff at the Center for Low Temperature Science, Tohoku University for providing assistance with the SQUID measurements. We also thank Messrs. M. Watahiki and S. Koyama for the overall assistance with the experiments, Dr. Y. Yamasaki for providing assistance with the fluorescence measurements, Dr. K. Iwasa for the warm encouragement and fruitful discussions, and Prof. R. Kadono, Drs. A. Koda, M. Miyazaki, and M. Hiraishi for supporting the preliminary heat capacity measurements at KEK. 
The X-ray fluorescence expereiments have been performed under approval of the Photon Factory Program Advisory Committee (Proposal No. 2009S2-008). 
This study was financially supported by Grants-in-Aid for Young Scientists (B) (22740209) and Scientific Researches (S) (21224008) from the MEXT of Japan. H. N. and Y. M. were financially supported by the Funding Program for World-leading Innovative R{\&}D in Science and Technology  (FIRST). 
\bibliography{LaCoMO3_4_arXiv}

\begin{thebibliography}{44}
\bibitem[{com()}]{comm_1}
\bibinfo{note}{The spin-state polarons are classified into magnetic polarons
  with spontaneous magnetization, not into normal polarons with electric
  polarization.}
\end{thebibliography}

\end{document}